\documentclass[conference,letterpaper]{IEEEtran}
\usepackage[utf8]{inputenc}
\usepackage[warn]{textcomp}
\usepackage{xcolor}
\usepackage{graphicx}
\usepackage[bookmarks=false,draft]{hyperref}
\usepackage{wrapfig}
\usepackage{ragged2e}
\usepackage{lipsum}  
\usepackage{stfloats}
\fnbelowfloat
\usepackage[bottom,multiple]{footmisc}
\usepackage{etoolbox}

\ifCLASSOPTIONcompsoc
  \usepackage[nocompress]{cite}
\else
  \usepackage{cite}
\fi

\ifCLASSINFOpdf
\else
\fi

\makeatletter
\patchcmd{\@maketitle}
  {\addvspace{0.5\baselineskip}\egroup}
  {\addvspace{-1\baselineskip}\egroup}
  {}
  {}
\makeatother

\usepackage[pscoord]{eso-pic}

\newcommand{\placetextbox}[3]{
  \setbox0=\hbox{#3}
  \AddToShipoutPictureFG*{
    \put(\LenToUnit{#1\paperwidth},\LenToUnit{#2\paperheight}){\vtop{{\null}\makebox[0pt][c]{#3}}}%
  }%
}%

\begin{document}

\placetextbox{0.5}{0.99}{\parbox{0.92\textwidth}{ \scriptsize{
    © 2018 IEEE. Personal use of this material is permitted. Permission from IEEE must be obtained for all other uses, in any current or future media, including reprinting/republishing this material for advertising or promotional purposes, creating new collective works, for resale or redistribution to servers or lists, or reuse of any copyrighted component of this work in other works. \\ \\
    \textbf{This is the accepted manuscript for SOFTCOM 2018. The final publication is available at IEEE Xplore via:} \url{http://dx.doi.org/10.23919/SOFTCOM.2018.8555785}
}}}

\title{KISS methodologies for network\\management and anomaly detection\\[1ex] 
  {\normalfont\large 
    \textsuperscript{\textleaf}Carlos Vega\IEEEauthorrefmark{1}\IEEEauthorrefmark{2}, Javier Aracil\IEEEauthorrefmark{1}\IEEEauthorrefmark{2} and Eduardo Magaña\IEEEauthorrefmark{1}\IEEEauthorrefmark{3} %
  }\\[-1ex]
}

\author{
\IEEEauthorblockA{\IEEEauthorrefmark{1}Naudit HPCN,\\ Parque Científico de Madrid, \\ \{carlos.vega, javier.aracil\\,eduardo.magana\}@naudit.es}
\and
\IEEEauthorblockA{\IEEEauthorrefmark{2} Escuela Politécnica Superior\\ Universidad Autónoma de Madrid\\
carlosgonzalo.vega@predoc.uam.es, \\ javier.aracil@uam.es} 
\and
\IEEEauthorblockA{\IEEEauthorrefmark{3}Departamento de Automática y Computación\\ Universidad Pública de Navarra\\
Campus Arrosadia, 31006, Pamplona \\
eduardo.magana@unavarra.es}
}


\renewcommand\IEEEkeywordsname{Keywords}
\markboth{}%
{Vega \MakeLowercase{\textit{et al.}}: KISS methodologies for network management and anomaly detection}


\IEEEtitleabstractindextext{%
\begin{abstract}
Current networks are increasingly growing in size and complexity and so is the amount of monitoring data that they produce. As a result, network monitoring systems have to scale accordingly. As a possible approach, horizontal scalability can be achieved by large data centralization systems based on clusters, which are expensive and difficult to deploy in a real production scenario. In this paper we propose and evaluate a series of methodologies, deployed in real industrial production environments, for network monitoring and management, from the architecture design to the visualization system as well as for the anomaly detection methodologies, that intend to squeeze the vertical resources and overcome the difficulties of data collection and centralization. \looseness=-1 
\end{abstract}


}

\maketitle
\IEEEdisplaynontitleabstractindextext
\IEEEpeerreviewmaketitle

\section{Introduction}\label{sec:introduction}

\IEEEPARstart{T}{he} information era has transformed the way societies communicate in a way not seen since the Gutenberg printer revolution, enabling the world to transmit information world wide almost instantaneously through the interconnection of computers, which today shape the Internet into a set of heterogeneous networks. Gone are the days in which BBS, Usenet, and e-mail were the only applications to be found in the Internet. As the number of users and devices joined the Internet, the value of the network outweighed the cost of belonging to it~\cite{metcalfe,metcalfe_web} and the Internet became omnipresent. 

Today, the ever growing number of users and services connected to the Internet~\cite{cisco_forecast_intro} poses new challenges for the network management systems, with more complex and heterogeneous systems interconnected by diverse protocols, offering services that are becoming more and more indispensable (e.g. shopping, social networks, etc.). This growing need for service ubiquity impacts the size of data centers~\cite{USAeReport} as well as the volume of data they manage~\cite{oecd,cisco_forecast}, which in turn affects the monitoring systems. \looseness=-1 

Monitoring systems are vital for Quality of Service (QoS) management and troubleshooting of incidences, therefore, a well suited system helps to prevent and locate the source of the issues, reducing the operational expenditure (OPEX) of network infrastructures. 

Additionally, QoS of most of these services has an strong relationship with the Quality of Experience (QoE)~\cite{qoeqos} which makes these services more sensible as they require greater availability. Service outages are one of the main concerns of any data center and network manager, as the cost per minute of data center outage can reach 8,000 USD per minute~\cite{Ponemon2016}, directly affecting OPEX. To prevent such incidents, traffic analysis is a crucial task to proactively identify potential sources of trouble, before they happen. \looseness=-1

In this paper we propose a series of \emph{Keep It Simple and Straightforward} (KISS) methodologies for anomaly detection in big data centers with limited resources, or, in other words, keeping a good balance between the size of the data center and the monitoring system, but without disregarding the effectiveness of the system. In particular, we explore the different stages involved in the monitoring process (collection, processing, anomaly detection, etc.) and propose simpler, yet effective, techniques.

Such techniques must be simple, in order to cope with the growing volumes of data, but also versatile to deal with the vagaries of the analysts. Exceptionally, we were privileged to deploy the researched and developed systems in real environments, in particular, a Spanish logistics company, in which we obtained invaluable feedback, that resulted extremely helpful for the development of the systems described below.

When a disruption arises in a particular QoS metric from a component of the data center, the impact may extend to the QoE, affecting the final user, either delaying the service process, or, in the most severe cases, causing a service outage. Locating the source of trouble in heterogeneous networks is not a trivial task, and it is necessary to distinguish between causes and effects, since a troubled component may be affected by another malfunctioning component it depends upon. \looseness=-1 

We will later discuss the system aspects, delving deeper into the proposed proactive and reactive methodologies for anomaly detection. \looseness=-1 

In what follows, we will provide insights into some alternatives from the literature and the current approaches used by large corporations. Afterwards, we will depict the network management architecture currently in production in the aforementioned infrastructure. Then, we will describe the different methods used for an effective monitoring which help to prevent and locate sources of trouble, specifically two pair of proactive/reactive methodologies for anomalous behaviour in time series and Web access events. We will also address the performance issues of these techniques. \looseness=-1 

\subsection{State of the art}

Monitoring systems present an ETL-like (Extract Transform and Load) cycle, comprising generation of management information, data collection, processing, storage, and the further analysis of the retrieved information and its visualization. 

\subsubsection{Event and traffic monitoring systems}

The first challenge of an effective monitoring system is the \textbf{collection of all relevant data from the network nodes}. These nodes produce data of different nature such as network traffic from the served applications (e.g. HTTP, Radius, etc.), log messages from the running services (e.g. Java exceptions), and management data from the different network nodes (e.g. CPU and memory usage). This raw information can be stored in high-speed storage for further processing and analysis.

During the last years, the interest about this type of anomaly detection systems has increased with the growth of data centers and the number of hosted services. However, most of these systems require either complex machine learning techniques, or deployment of agents in the supervised machines. 

For instance, the system proposed by Zhang et al.~\cite{TaskInsight} seeks to find the root cause of the incidences and anomalies through the behavioral analysis of the threads and resources used by executed applications. To obtain this information, the deployment of Python agents in different machines is a prerequisite, with the subsequent difficulties from the constraints of production systems. 

Bahl et al. also propose a solution that uses agents to monitor network packets sent and received by hosts~\cite{sherlock}. Although this solution finds performance problems \textit{``without requiring modifications to existing
applications and network components''}, it requires changes in production systems which often present different operating systems or environments.

As stated above, \textbf{the proposed system must be as omniscient as possible}, gathering as much information as possible, but it is also important to know that monitoring systems cannot be omnipotent, as some solutions seem to assume. This is a hurdle to overcome, in which large infrastructures are often organized in different and outsourced departments, with strong constraints regarding security, confidentiality and availability, disallowing any process that may interfere the proper behavior of their systems. Consequently, a passive, centralization system should be in order, contrary to the aforementioned systems approaches. These passive systems also face security obstacles disseminated across the network such as firewalls or intrusion detectors that require consent from managers.

\subsubsection{Data centralization}

Many applications already provide event records about resource and traffic usage with high level of detail. Generally the challenge is not to create monitoring information but to collect it for further analysis, that otherwise would stay in the machine waiting for manual inspection from the system managers seeking answers for troubleshooting.

Big data solutions employ proportionless systems with excessive resources (e.g. a cluster of nodes) which feels \textit{de~trop} in infrastructures where resources are scarce. For example, Kepner et al. compared different database technologies such as Cassandra, Oracle and Accumulo, with the latter offering the best performance. Even though they achieve 115~million inserts per second in an Accumulo database, they require a vast amount of resources, specifically, 216 Accumulo nodes and 1,296~ingest processes, with an average performance of 100,000~entries/second per ingest process~\cite{achieving}.

Hence, \textbf{monitoring systems must be dimensioned and designed in accordance with the data center size}, squeezing vertical resources and taking advantage of its privileged standpoint in the enterprise network, without disregarding performance.

\subsubsection{Information processing}

In fact, the third step is to process all this information by different means, through high-performance traffic dissectors~\cite{NDPI,HTTPD} or log-parsing engines such as \textit{FluentD} or \textit{Logstash}. Afterwards, the processed data can be stored in non-relational databases like \textit{Elasticsearch}, \textit{Solr} or \textit{Splunk}. Is not the purpose of this work to review all these alternatives but the performance must be in accordance with the rate of collected information. As we process the raw information to get enriched records we tend to aggregate the information and reduce its amount, reducing the need of performance and leaving space for versatility.

\subsubsection{Anomaly detection}
\label{sec:anodetec}

Last but not least, anomaly detection systems must as well fit the above-mentioned needs of versatility and performance. There are complex solutions with strong mathematical background~\cite{quantile,dictyogram,unsupervised_outlier,mining_outliers} able to detect subtle deviations in the metric behavior. Notwithstanding, is essential to dispose of a simpler and faster system able to detect anomalies among the numerous monitored metrics and their corresponding time-series in a timely manner.

Furthermore, the human element calls for alarm containment, since recurrent alerts, whether they are legitimate or not, produce apathy in the system managers, who will mute future warnings, loosing confidence in the alarm system. In other words, it is preferable to generate alarms about severe incidents with probable impact at the expense of muting minor deviations, unless the latter happens repeatedly over time, which requires adjusting the intervals between the first alert and the successive warnings.

Systems such as the proposed by Calheiros et al.~\cite{iforest_anomaly} use \textit{Isolation Forest}~\cite{iforest}, proposed by Liu et al., to detect anomalies. This method benefits from the fact that anomalies are scarce and usually isolated from the normal observations. They randomly choose the characteristics to be considered and split them recursively creating a series of partitions. The method isolates the anomalous values near the root of binary trees, using them to represent the available data. \looseness=-1 

However, in their own words, \textit{``Initial attempts of applying iForest with such small number of attributes did not lead to good anomaly detection power on the approach.''}. Thus, the authors obtained a new group of attributes from previous observations, generating a new set of \textit{expected values} that define the \textit{usual behaviour}. To do so, they take advantage of weekly and hourly behaviour cycles of the data, storing the median of the values from each cycle, that define the \textit{expected values} of the time series. We will compare our proposed technique to this system in section~\ref{sec:evaluation}.

\section{Architecture of the system}

\begin{figure}[t]
\begin{center}
\includegraphics[width=\columnwidth]{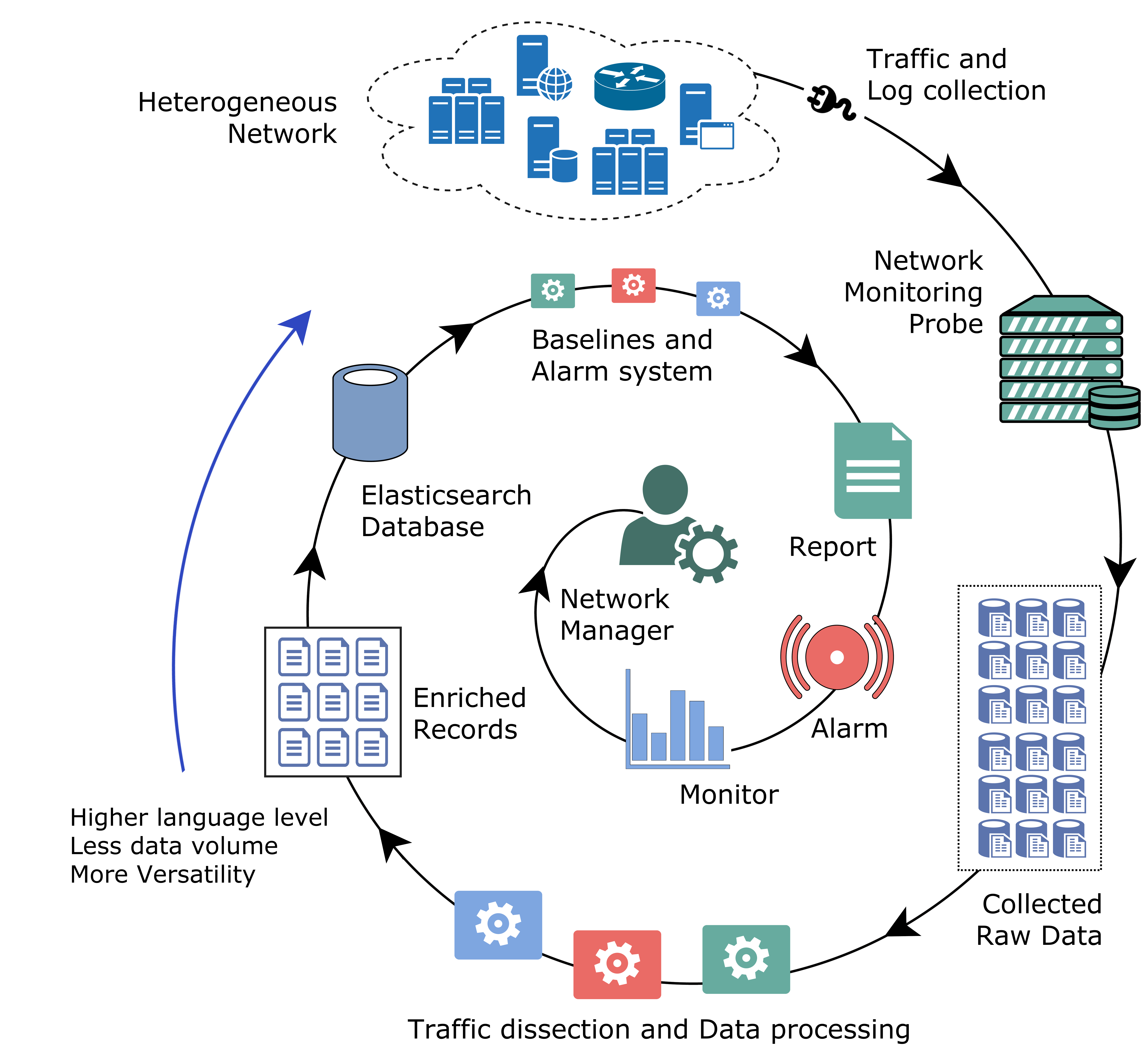}
\vspace{-1.7em}
\caption{Stages involved in the monitoring process. As we climb from the wire to the charts the information is aggregated, from raw data to enriched records and finally the charts, notifications or reports. The amount of data is reduced at each stage, as well as the performance requirements while versatility requirements increase.}
\label{fig:spiral}
\vspace{-2em}
\end{center}
\end{figure}

We now dwell on the details of the proposed architecture as well as the mechanisms behind the anomaly detection system, which is the monitoring system core. We will describe two different examples of techniques. The proposed architecture is depicted in Figure~\ref{fig:spiral}.

\subsection{Two phase monitoring system}

Locating root causes may be a huge challenge for an automatic monitoring system, hence, \textbf{we propose a two phase system of detection and inspection}. Raw network traffic at multi-Gb/s speeds~\cite{M3OMON} and raw event logs are centralized~\cite{loginson} in a high-end network probe. Concurrently, a summary, in the form of enriched records, is stored for real-time monitoring in a second lower-end server
, featuring visualization dashboards for the analysts and service managers. These enriched records provide information about connections, sessions or transactions which have higher context that packet-level information. As Goodall et al. state, \textit{``analysts often lose sight of the “big picture” while examining these low-level details''}\cite{goodall}, referring to packet-level analysis.

\textbf{Traffic and log collection is done in a centralized fashion}, gathering the traffic from a Switched Port Analyzer (SPAN) port and the log information via standard applications, available in most systems, such as syslog, SNMP or ssh. These methods avoid disrupting production systems as much as possible. When a disruption or anomalous behaviour is detected on the summarized data, the analyst can perform further inspection in the raw data.

\subsubsection{Network Probe for collection}

The deployed network probe is an Intel\textregistered~Xeon\textregistered~CPU E5-2640~v2 @~2.60GHz system with 8 cores, 64~GB of RAM, and four 15TB RAID~0 storage systems, to store, in three of them, traffic captured from different network interfaces. Each interface captures different subnets or VLANs. The remaining RAID is used for inspection analysis, data processing and other purposes.

The lessons learned during the development of the log centralization system Loginson~\cite{loginson} have been applied for the development of the system deployed in the aforementioned network. M3OMON and HPCAP Intel NIC driver~\cite{M3OMON} are used for traffic collection and generation of flow information from the traffic. Collection of log messages is done through direct copy at particular intervals while component information (e.g. memory usage) is retrieved through SNMP. 

Collected information is processed according to the analysis requirements. Usually, the huge amount of raw data calls for low-level high-performance solutions to dissect the different traffic layers and protocols, such as HTTPanalyzer~\cite{HTTPD}.

\subsubsection{Virtual Machine for summarization}

In 2015, the deployed VM used modest resources, specifically, 8GB of RAM, 100GB of storage and four processing cores. As 2017 the VM has doubled the storage and now features 12~GB of RAM and eight processing cores. To store the summarized data, we chose \textit{Elasticsearch} due to its performance, and analytical features which assist the statistical calculations of the analysis (queries to calculate aggregations, statistics such as standard deviations, medians, etc.). While traditional databases provide diverse and refined query features, our use case, sacrifices these characteristics in favor of speed, analytical features.

The internal storage architecture of the Elasticsearch database, stores the documents in specific {\em indices}, made of a group of {\em shards}, which are essentially {\em Lucene Inverted Indices}. The different operations (aggregation, filter, etc.) are executed at shard level, taking advantage of this given parallelism. 

Once the raw data is processed in the network probe, the corresponding enriched records are indexed in the VM, using an Elasticsearch importer\footnote{Similar to: \url{https://github.com/carlosvega/ElasticsearchImporter}}. Tools for baseline generation, anomaly detection and others have been developed in Python, using the high-level \textit{elasticsearch-dsl} library. For data visualization, we use different technologies such as \textit{Kibana} (vis. plugin for Elasticsearch) or \textit{Plotly}. \looseness=-1 

This simple architecture makes complete use of the vertical resources, requiring fewer machines to work. This way, availability can be improved through replication active-active or pasive-active of the monitoring system. Also, its extent can be broaden through replication of the same architecture in different departments. Also, elasticsearch database allows for distribution and clusterization of multiple nodes. 

\subsection{Time Series Monitoring}

In the next sections we describe two examples of monitoring and anomaly detection with two pairs of proactive/reactive methodologies. Specifically, we now dwell on the detection of disruptions or anomalies in time series as well as the creation of baselines to assist this process.\looseness=-1 

\subsubsection{Proactive methodology I: Baseline creation}

\begin{figure}[b!]
\vspace{-1em}
\begin{center}
\includegraphics[width=\columnwidth]{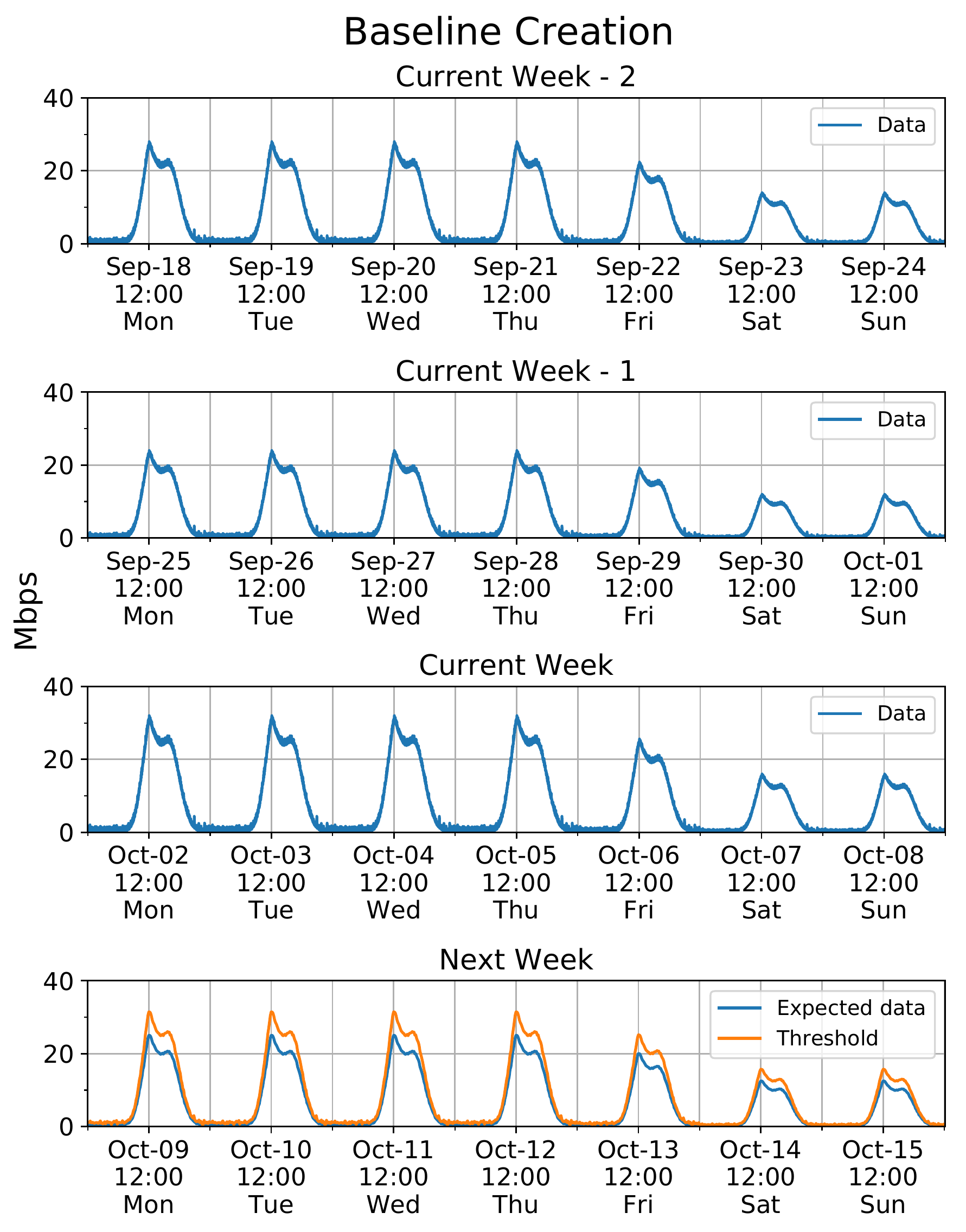}
\vspace{-2em}
\caption{Baseline creation process making use of previous data.}
\label{fig:traffic_weeks}
\end{center}
\vspace{-1em}
\end{figure}

\begin{figure}[b!]
\begin{center}
\vspace{-0.5em}
\includegraphics[width=\columnwidth]{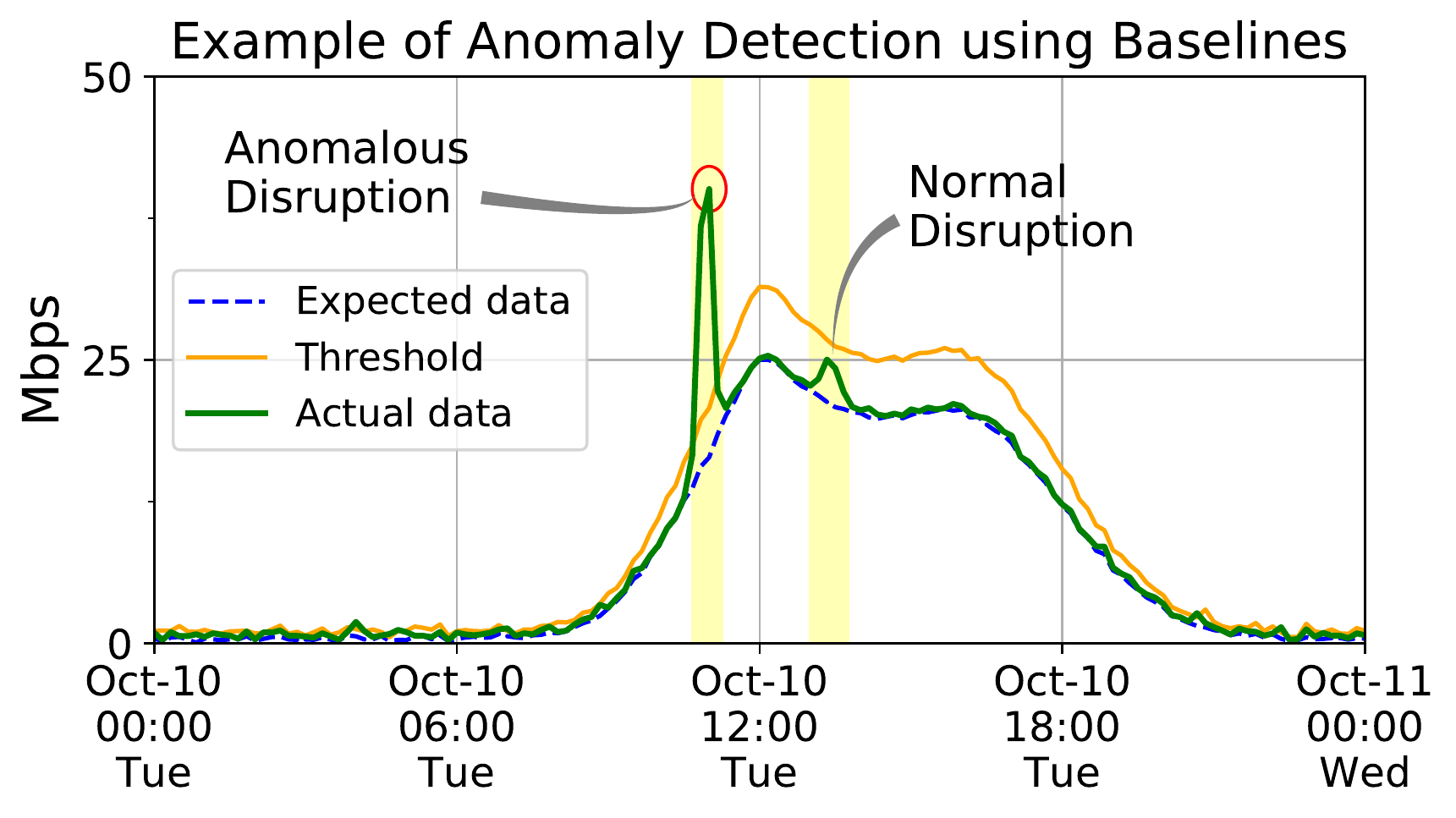}
\vspace{-2em}
\caption{Anomaly detection using the calculated baseline.}
\label{fig:traffic_anomaly}
\end{center}
\end{figure}

With a big enough historical record of data, we can conduct proactive analysis for the generation of baselines that define the usual behaviour of the analyzed metrics. To do so we consider the previous \textbf{N}~weeks. The larger the number, the more rigid the baseline will be, making it harder to adapt to a new trend, should it appear. As seen, during the weekend the traffic is lower than on working days. A moving window of size~\textbf{W} (e.g. 10 min.) is considered, then, the data is grouped by \underline{\smash{weekday, hour and minutes}}, rounded up to the window size. For each of these groups, which consists of 4x10 values (or \textbf{NxW} values), the median is calculated, yielding a time series (here, 10 min. resolution) of expected values for the next week.

For a more detailed example, let us consider the time series~$\{X_{n}, n=1,\ldots\}$ which represents the collected traffic samples at a given time $n$, specifically in Fig.~\ref{fig:traffic_weeks} the current one and the 2 previous weeks with, one data point per minute. Let the resulting moving-averaged time series be represented by $\{\hat{X_{n}}, n=1,\ldots\}$. In order to predict the time series value at a given time $k$ we consider the random vector \looseness=-1 
\begin{equation} \label{eq:mvag}
\mathcal{X}_m^T(k)=(\hat{X_{k-mT}}, \hat{X_{k-(m-1)T}}, \ldots, \hat{X_{k-T}})
\end{equation}

whereby $T$ denotes the time a week before the current time $k$ and $m$ is the estimation span in number of weeks. Basically, the vector is made up with the past values of the time series, $1,\ldots, m$ weeks before the current time. Let $f_m(k)$ be the sampling distribution of $\mathcal{X}_m^T(k)$, as defined in equation~\ref{eq:mvag}. 

Thus, the time series $\{\hat{X_{n}}, n=1,\ldots\}$ forecast at time $k$ is given by $median[f_m(k)]$. The median is a robust estimator with a breakdown point of 50\%. Besides, another similar series with the standard deviation of the data groups instead of the median is calculated. Together they set a threshold for each moment of the week, as seen in the bottom chart of Figure~\ref{fig:traffic_weeks}. 

This baseline will assist reactive tasks such as manual visual inspection of charts or alarm systems for anomaly detection.

\subsubsection{Reactive methodology I: Anomaly detection}

Once the baseline is calculated, we can detect disruptions in time series with values higher than their threshold occurred over a given time of grace. The tool uses static thresholds, or functions of values. We can then use the baseline as well as other related indicators for increased effectiveness. For instance, we can alert if the incoming number of flows is higher than the threshold and at the same time the number of outgoing flows is zero, since this scenario might represent a service outage.

This kind of alarms provides versatility in three different aspects. First, it allows the selection of concrete time intervals and time aggregations of the selected data, as well as time of grace conditions. Secondly, it offers filtering flexibility for different granularities. Finally, ad-hoc functions to define conditions can be set in order to use other metrics.

Figure~\ref{fig:traffic_anomaly} shows two examples of abnormal behaviour of the metric. The first one, (red circle) occurs before noon and shows how the data value surpassed the expected value and the baseline threshold, during the time of grace interval. In this scenario a notification would be generated for the network manager. On the other hand, the second disruption, past noon, does not pass these bounds, and hence, no warning is needed. 

\subsection{IP Geo-location Monitoring}

We now introduce another example of simple methodologies for anomaly detection, focused on web visits behaviour.

\subsubsection{Proactive methodology II: Web access report}

Web analytics have an important role in market research and security management. Geo-location of visitors helps to define access patterns and product strategies.

The Regional Internet Registries databases such as the ARIN or APNIC, among others, are the primary source of tables relating geographical locations to IP address ranges. Services like \textbf{ip2location} offer relational databases with different geographical resolutions. The proposed method is an extension of a previously developed tool~\footnote{\url{https://carlosvega.github.io/elasticGeoIPMaps/}} for data visualization of IP-location in off-line environments in the form of a map. \looseness=-1  

\begin{figure*}[t!]
\vspace{-1em}
\begin{center}
\includegraphics[width=\textwidth]{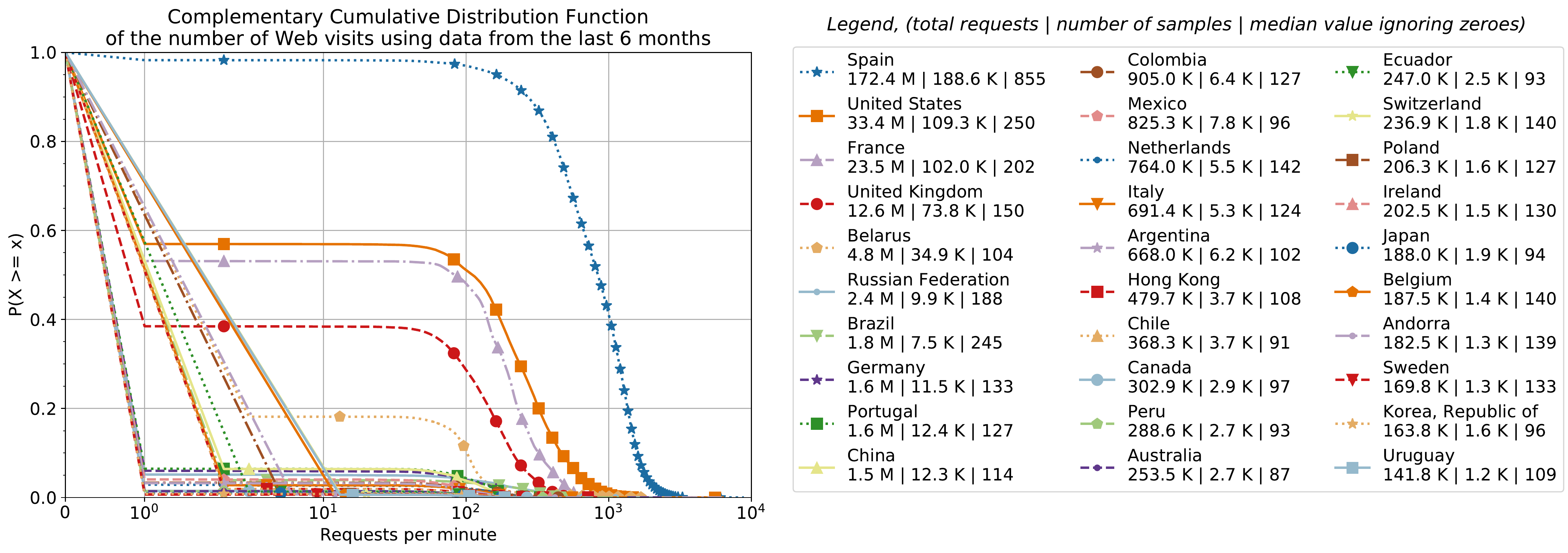}
\vspace{-2em}
\caption{CCDF of the requests per minute per country. Uncommon origins have very low probability of occurrence. X axis is in symlog scale}
\label{fig:ccdf_countries}
\end{center}
\vspace{-1.5em}
\end{figure*}

\subsubsection{Reactive methodology II: Anomalous visits detection}

Anomalous visits can hide potential attacks which can undermine the proper operation of the services and affect their QoS. Detecting these threats is crucial to prevent service outages and roll out countermeasures. In the early stage of an attack, this abnormal behavior, often consisting of visits from uncommon places, has little impact in terms of Web visits, and may also be legitimate, but the number can rapidly increase in case of a real attack is happening.

Modeling the behavior of countries with few visits is challenging, since in harmless scenarios these countries show little activity or infrequent small bursts of visits, if any. Figure~\ref{fig:ccdf_countries} shows how uncommon origins have low probability of occurrence in the first place. In this figure, Spain, United States, France and United Kingdom are the four most common sources of Web visits. Belarus also stands out, since, a single IP address was the source of most visits from this country, turning out to be a shipment tracking website. 

However, the repeated observation of this anomalous activity, usually higher than the previously observed behavior, uninterrupted during a considerable amount of time is what yields an alarm notification, sent to the manager in charge for further analysis. For example, checking if the source IP addresses are in any Threat Intelligence Platform (TIP) list 
or inspecting the behaviour of these IPs in depth. 

\section{System Evaluation}
\label{sec:evaluation}

In this section we address the performance details of the proposed methodologies, from the architecture above described to the proactive and reactive systems developed in accordance with the aforementioned requirements of performance, simplicity and versatility.

\subsection{Baseline and Anomaly Detection Performance}
Currently, on a weekly basis, the baseline generation system calculates about 600~baseline time series using the 8~previous weeks of data (with 1 minute resolution), considering more than 48~million data points. These baselines correspond to diverse metrics from different monitored services. The baseline generation takes 1 hour and 15 minutes to complete. As explained, these baselines are used by both the alarm system and manual inspection of the charts by network managers.

The alarm system checks the most recent data (30 to 60 minutes of previous data), from more than 300~time series, seeking for anomalies. This process takes less than 20~seconds in normal conditions and always less than a minute during high load periods. For the best-case the speed is around 1,000~elements per second. This process is done in parallel with the data indexation, as well as the regular inspection of dashboards by managers and analysts which generate queries to the database. Today more than 10 million documents per day are indexed in the database and is expected to grow.

The system proposed by Calheiros et al.~\cite{iforest_anomaly} seen in Section~\ref{sec:anodetec}, uses aggregated data in 30 minute intervals. The performance observed by Calheiros et al for calculating their \textit{anomaly score} was the following: \textit{``calculation of anomaly scores of 2,722 data points […] took 28 milliseconds […] The training time was in 579 milliseconds...''}. This study doesn't consider the time it takes to query data stored in the database with continuous queries from other systems. It neither considers how long it takes to calculate what they call \textit{derived attributes}.

Last but not least, they do not consider the fact that usually, the metrics of interest are stored together with other relevant information such as the service name, and other metadata which also makes the indexed elements bigger.\looseness=-1  

\subsection{Performance of the Geo-location Monitoring}

The map system considers the top 100,000 IP addresses by number of visits from the chosen time interval and features a per-country tooltip with the top 10 IPs with more requests. 

This solution is able to generate an interactive map with the last week data in less than a minute. The amount of data considered for the four monitored web services is around 400,000 elements per week. Hence, in the worst case, it could be up to 100,800 different IPs for an specific service. The performance of the anomalous visits detection system is similar to the anomaly detection system aforementioned. 

\begin{figure}[b!]
\vspace{-1em}
    \centering
 \includegraphics[width=\columnwidth]{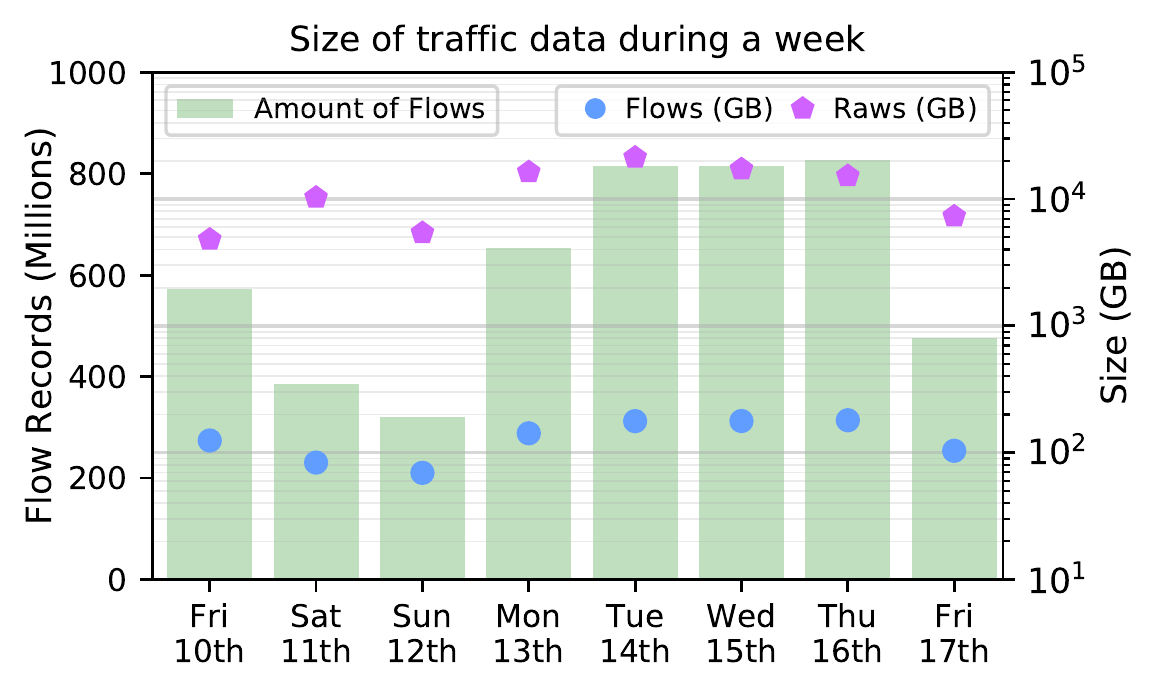}
    \caption{Sizes of raw traffic data from a Spanish logistics company, and their corresponding amount of flow records, for each weekday.}
    \label{fig:raws_flows_week}
\end{figure}

\subsection{Extent of the Monitoring System}

Recently, the network coverage of the monitoring system deployed in a Spanish logistics company has expanded. Figure~\ref{fig:raws_flows_week} shows the amount of network traffic captured per day during a particular week, as well as the amount and size of TCP flows. In aggregate terms, the network probe captures more than 10~Gbps of traffic, with more than 2.5~million packets/s, from 1.5 million concurrent connections.

Regarding the summarized data, the amount of indexed documents in the Elasticsearch database doesn't cease to grow. For example, in February 2017 the number of indexed documents per day was 2~million docs/day, and by January 2018 this number has already surpassed the 10 million docs/day and more than 20~GB/month. A remarkable number considering the number of concurrent tasks to be done, such as dashboard visualization and data query, as well as the real-time alarm systems and proactive analysis tools (e.g. baseline system). \looseness=-1

\section{Conclusions}

Traditional approaches for network management over-provision resources, wasting vertical scalability through large cluster centralization systems. They also assume absolute control over the monitored nodes, proposing agent-based solutions, disregarding the bureaucratic hindrances that impede deployments in production systems. An effective network management must not disrupt the normal operation of the center. It must also be proportionate to the size of the monitored network, optimizing vertical scalability. \looseness=-1  

In this paper we proposed simple techniques to address the aforementioned needs. Specifically, the architecture design of the system, data centralization systems, and data analysis methodologies. We evaluated the solution in a real enterprise environment, with wide monitoring coverage, and achieved great performance through vertical utilization of resources. Additionally, we contributed releasing proof-of-concept versions of the tools used in our production systems.

\ifCLASSOPTIONcompsoc
  \section*{Acknowledgments}
\else
  \section*{Acknowledgment}
\fi

The authors would like to thank MINECO, received through grant TEC2015-69417 (TRAFICA)

\ifCLASSOPTIONcaptionsoff
  \newpage
\fi

\bibliographystyle{IEEEtran}
\bibliography{IEEEabrv,kiss}

\end{document}